\documentclass{article}
\usepackage{amsmath,amssymb,amsfonts,amsthm}
\usepackage[russian]{babel}
\usepackage{color}
\usepackage{graphicx}
\usepackage{wrapfig}

\begin{document}

\flushbottom

\numberwithin{equation}{section}

\renewcommand{\figurename}{Fig.}
\def\refname{References}
\def\proofname{Proof}

\newtheorem{theorem}{Theorem}
\newtheorem{propos}{Proposition}
\newtheorem{remark}{Remark}

\def\tens#1{\ensuremath{\mathsf{#1}}}

\if@mathematic
   \def\vec#1{\ensuremath{\mathchoice
                     {\mbox{\boldmath$\displaystyle\mathbf{#1}$}}
                     {\mbox{\boldmath$\textstyle\mathbf{#1}$}}
                     {\mbox{\boldmath$\scriptstyle\mathbf{#1}$}}
                     {\mbox{\boldmath$\scriptscriptstyle\mathbf{#1}$}}}}
\else
   \def\vec#1{\ensuremath{\mathchoice
                     {\mbox{\boldmath$\displaystyle\mathbf{#1}$}}
                     {\mbox{\boldmath$\textstyle\mathbf{#1}$}}
                     {\mbox{\boldmath$\scriptstyle\mathbf{#1}$}}
                     {\mbox{\boldmath$\scriptscriptstyle\mathbf{#1}$}}}}
\fi

\newcommand{\diag}{\mathop{\rm diag}\nolimits}
\newcommand{\const}{\rm const}
\def\div{\operatorname{div}}

\begin{center}
{\Large\bf The Hess\,--\,Appelrot system\\
and its nonholonomic analogs\\}

\bigskip

{\large\bf Ivan~A.\,Bizyaev$^1$,
Alexey~V.\,Borisov$^2$,
Ivan~S.\,Mamaev$^3$\\}
\end{center}

\begin{quote}
\begin{small}
\noindent
$^{1,2,3}$ Steklov Mathematical Institute, Russian Academy of Sciences,\\
ul.~Gubkina~8, Moscow, 119991 Russia\\[2mm]
$^1$ E-mail: bizaev\_90@mail.ru\\
$^2$ E-mail: borisov@rcd.ru\\
$^3$ E-mail: mamaev@rcd.ru
\end{small}

\bigskip
\bigskip

\begin{small}
\textbf{Abstract.} This paper is concerned with the nonholonomic Suslov
problem and its generalization proposed by Chaplygin. The issue of the
existence of an invariant measure with singular density (having
singularities at some points of phase space) is discussed.

\smallskip


\smallskip

\end{small}
\end{quote}

\footnotetext[0]{This work was supported by the Russian Scientific Foundation
(project No. 14-50-00005).}

\section{Introduction}

The Suslov problem is one of the model problems in nonholonomic mechanics, which describes the motion of a rigid body
with a fixed point whose projection of angular velocity onto the body-fixed axis
is zero (left-invariant nonholonomic constraint).

In~\cite{bkm, Kozlov, Tatarinov1988, Bloch2014, Vagner41, bkm1}, a qualitative analysis is made of
the dynamics of the Suslov problem.
In particular, cases of the existence of an invariant measure and additional first integrals are presented
and the topological type of integral manifolds is investigated.
We note that in integrable cases the two-dimensional integral manifolds can be different from tori (which are described
by the Euler\,--\,Jacobi theorem and occur, for example, in the Chaplygin problem of the rolling motion of a dynamically asymmetric
ball \cite{Kozlov}).

The Suslov problem is closely related to another nonholonomic system, a~Chaplygin sleigh \cite{BPMM}. The latter can be obtained
from the Suslov problem by contracting the group $SO(3)$ to the group $E(3)$~\cite{Kozlov2015}. Thus,
the Suslov problem is a compact version of the Chaplygin sleigh problem.
Of interest is the dynamics of the Suslov\,--\,Chaplygin system on the group $SO(2,1)$, which can obviously have
compact and noncompact trajectories.

The equations of motion in the Suslov problem with a nonzero gravitational field are nonholonomic analogs of the Euler-Poisson
equations, which in the general case possess no smooth invariant measure,
but, depending on the system parameters, can admit both regular and chaotic behavior.
Chaotic dynamics and reversal phenomena in the Suslov problem are examined in~\cite{bkm}.

In this paper we consider the case in which the system is not integrable by quadratures by the Euler\,--\,Jacobi theorem,
but its behavior is regular. In this case, this system admits $n-2$ first
integrals ($n$ being the dimension of the phase), which can be used to reduce the problem to analysis of the flow on a~two-dimensional
manifold.

We show an isomorphism between the case of regular dynamics in the Suslov
problem, which was found in~\cite{bkm}, and the classical Hess system in
the Euler\,--\,Poisson equations. In the Appendix to this paper we present the
most well-known facts about the Hess case and a critical analysis of the
recent\linebreak publications~\cite{Lubowiecki2012, Lubowiecki2012II, Belaev2015}.

The analogy between the Suslov problem and the Hess case is closely
related to the fact that the dynamics on an invariant manifold is
``dissipative'', i.e., it can possess attracting sets (see~\cite{Kozlov}).
We note that both problems reduce to solving the Riccati equation (for the
Hess case this result was obtained by P.\,A.~Nekrasov~\cite{Nekrasov}, and
for the Suslov problem by G.~Vagner~\cite{Vagner41}). Analysis of this
equation in the Suslov problem without a gravitational field yielded the
formula~\cite{Fedorov2009} for the scattering angle. It turned out that if
there is an additional  first integral, the rotation axis of the body
reverses direction. We note that this result is natural, and perform a
more detailed analysis of the scattering problem. This analysis is closely
related to the existence of smooth or analytic integrals~\cite{Knauf}.

We also discuss the problem of the existence of an invariant measure with singular density
(i.e., a measure having singularities
at some points of phase space, see also~\cite{Kozlov2016, Bizyaev2016}).
Such a measure can considerably influence the system dynamics, i.e., determine some singularities of asymptotic dynamics
and the related scattering problem.
We conclude by discussing a combination of the Suslov problem and the Chaplygin ball rolling problem.
We present a dynamical interpretation of this problem.
In this case, the closed system of equations for angular velocities does not decouple, but nevertheless the system possesses an
invariant measure with singular density. We formulate a number of new problems for the study of the dynamics of this system.

In recent works the classical Hess case is called the Hess-Appelrot system\footnote{Considering the name used for this case in
search systems in the Web
(see, e.g., the recent papers
\cite{Lubowiecki2012,Lubowiecki2012II,DR01,DR02}), we have decided to keep to the correct name, i.e., the Hess
case, throughout the paper and to call it the Hess-Appelrot case in the title of the paper.}. In fact,
G.Appelrot did not find this case due to errors in his calculations. The history of this issue is described in detail
in the Appendix, which is concerned with analysis of the Hess case.

\clearpage

\subsection{A singular measure and invariant manifolds}
\label{sec1.2}

We show that if the dynamical system
\begin{equation}
\label{eq11}
\dot {\boldsymbol x}=\boldsymbol v(\boldsymbol x)
\end{equation}
admits an invariant measure that is smooth almost everywhere, then the
points at which the measure has singularities form invariant sets of the
system (see~\cite{Kozlov2016}, Section 5~--- Invariant measures with
density of alternating signs).

\begin{propos}
Suppose that the system~\eqref{eq11} possesses a (smooth) invariant measure $\mu=\rho(\boldsymbol x)dx_1 \ldots dx_n$
whose density can vanish. Then the submanifold
$$
\mathcal{M}_0=\{\boldsymbol x|\rho(\boldsymbol x)=0\}
$$
is an invariant submanifold of the system.
\end{propos}

{\proof Write the Liouville equation for the density of the invariant measure in the form
\begin{equation}
\label{eq1}
\div \rho \boldsymbol v=\dot \rho+\rho \div \boldsymbol v=0, \quad \dot \rho=\sum_i\frac{\partial \rho}{\partial x_i}v_i(\boldsymbol x).
\end{equation}
It can be seen that on the submanifold $\mathcal{M}_0$ the derivative $\dot \rho=0$,
hence, this submanifold is invariant.}

The following proposition is proved in a similar way.
\begin{propos}
Suppose that the density of the invariant measure in the\linebreak
system~\eqref{eq11} can go to infinity in such a way that at these points
the function $g(x)=\frac{1}{\rho(x)}$ turns out to be smooth. Then the
submanifold
$$
\mathcal{M}_s=\{\boldsymbol x|g(\boldsymbol x)=0\}
$$
is invariant too.
\end{propos}

{\proof Rewrite the Liouville equation~\eqref{eq1} in terms of the function $g(\boldsymbol x)$:
$$
\dot g-g\div \boldsymbol v=0.
$$
This gives the conclusion of the proposition.}

\clearpage	
\section{The Suslov problem with a nonzero gravitational field}	
\subsection{Equations of motion}
\label{sec2.1}
Consider the motion of a rigid body with a fixed point subject to the nonholonomic constraint
\begin{equation}
(\boldsymbol{\omega}, \boldsymbol{e}) = 0,
\label{Eq:SuslovConstraint}
\end{equation}
where $\boldsymbol{\omega}$~is the angular velocity of the body and $\boldsymbol{e}$
is the body-fixed unit vector.
The constraint \eqref{Eq:SuslovConstraint} was proposed by G.\,K.\,Suslov in
\cite[p.~593]{Suslov46} (for the realization of this constraint, see~\cite{bkm1}).

To parameterize the configuration space, we choose a matrix of the direction cosines ${\bf Q}\in SO(3)$
whose columns are the unit vectors of a fixed coordinate system that are referred to a moving coordinate system
$Ox_1x_2x_3$ rigidly attached to the body:
$$
{\bf Q}=\begin{pmatrix}
\alpha_1 &  \beta_1 & \gamma_1    \\
\alpha_2 &  \beta_2 & \gamma_2  \\
\alpha_3 &  \beta_3 & \gamma_3
\end{pmatrix}\in SO(3).
$$

The equations of motion in the moving coordinate system in the case of potential external forces have the form
\begin{equation}
\label{Biz01}
\begin{gathered}
\small
{\bf I}\dot{\boldsymbol{\omega}}={\bf I}\boldsymbol{\omega}\times\boldsymbol{\omega} + \lambda\boldsymbol{e}+
\boldsymbol \alpha\times \frac{\partial U}{\partial \boldsymbol \alpha}+
\boldsymbol \beta\times \frac{\partial U}{\partial \boldsymbol \beta}+\boldsymbol{\gamma}\times\frac{\partial U}{\partial\boldsymbol{\gamma}}, \\
\lambda =-\frac{\left({\bf I}^{-1}\boldsymbol{e}, {\bf I}\boldsymbol{\omega}\times\boldsymbol{\omega} + \boldsymbol \alpha\times \frac{\partial U}{\partial \boldsymbol \alpha}+
\boldsymbol \beta\times \frac{\partial U}{\partial \boldsymbol \beta}+ \boldsymbol{\gamma}\times\frac{\partial U}{\partial\boldsymbol{\gamma}}
\right)}{({\bf I}^{-1}\boldsymbol{e}, \boldsymbol{e})},
\end{gathered}
\end{equation}
\begin{equation*}
\dot{\boldsymbol{\alpha}}=\boldsymbol{\alpha}\times\boldsymbol{\omega}, \quad \dot{\boldsymbol{\beta}}=\boldsymbol{\beta}\times\boldsymbol{\omega}, \quad
\dot{\boldsymbol{\gamma}}=\boldsymbol{\gamma}\times\boldsymbol{\omega},
\end{equation*}
where $\boldsymbol \alpha=(\alpha_1, \alpha_2, \alpha_3)$, $\boldsymbol \beta=(\beta_1, \beta_2, \beta_3)$,
$\boldsymbol \gamma=(\gamma_1, \gamma_2, \gamma_3)$, and $U$~is the potential energy of the external forces.

Let us choose the moving body-fixed coordinate system $Ox_1x_2x_3$ in such a way that $Ox_3 \| \boldsymbol{e}$ and the axes
$Ox_1$ and $Ox_2$ are directed so that one of the components of the inertia tensor of the body vanishes: $I_{12}=0$.
In this case, the constraint equation \eqref{Eq:SuslovConstraint} and the tensor of inertia ${\bf I}$
of the rigid body can be represented as
\begin{equation}
\begin{gathered}
\label{Biz03}
\omega_3=0, \quad
{\bf I}=\begin{pmatrix}
I_{11} &  0 & I_{13}    \\
0 &  I_{22} & I_{23}  \\
I_{13} &  I_{23} & I_{33}
\end{pmatrix}.
\end{gathered}
\end{equation}

We shall also assume that the body moves in a gravitational field, so that the potential energy is
$$
U=(\boldsymbol b, \boldsymbol \gamma), \quad \boldsymbol b=-mg \boldsymbol r,
$$
where $\boldsymbol r$~is the body-fixed radius vector of the center of
mass, $m$~is the mass of the body, and $g$~is the free-fall acceleration.
In this case, in the system~\eqref{Biz01} the equations governing the
evolution of $\boldsymbol \omega$ and $\boldsymbol \gamma$ decouple. Using
the constraint~\eqref{Biz03}, we can represent this system as
\begin{equation}
\begin{gathered}
\label{Biz131}
I_{11}\dot{\omega}_1=-\omega_2(I_{13}\omega_1 + I_{23}\omega_2) + b_3\gamma_2-b_2\gamma_3, \\
I_{22}\dot{\omega}_2=\omega_1(I_{13}\omega_1 + I_{23}\omega_2) + b_1\gamma_3-b_3\gamma_1, \\
\dot{\gamma}_1=-\gamma_3\omega_2, \quad \dot{\gamma}_2=\gamma_3\omega_1, \quad \dot{\gamma}_3=\gamma_1\omega_2 - \gamma_2\omega_1.
\end{gathered}
\end{equation}

The system~\eqref{Biz131} possesses an energy integral and a geometrical integral:
\begin{equation}
\label{Biz11}
E=\frac{1}{2}(I_{11}\omega_1^2 + I_{22}\omega_2^2) + (\boldsymbol b, \boldsymbol \gamma), \quad F_1=\boldsymbol{\gamma}^2=1.
\end{equation}

In the general case, for the system~\eqref{Biz131} to be integrable by the Euler\,--\,Jacobi theorem, one needs an additional
integral $F_2$ and a smooth invariant measure~\cite{Kozlov, BolsinovHam}.

This system is a nonholonomic analog of the classical Euler-Poisson
equations describing the dynamics of a rigid body with a fixed point in a gravitational field
(see~\cite{dtt} and references therein). As is well known (see \cite{dtt}), the Euler-Poisson equations
possess an area integral, a standard invariant measure and a Poisson structure (given by the algebra $e(3)$).
In the general case, these objects are absent for the system \eqref{Biz131}, but can exist under certain restrictions on the
parameters.
Other examples of nonholonomic systems exhibiting chaotic behavior are given
in \cite{Sat, Jalnine}.

A special feature of the system \eqref{Biz131} is that there can exist not only an invariant measure with everywhere smooth positive
density, but also a singular measure whose density has singularities on some invariant manifolds of the system.
For the system \eqref{Biz131} we first list cases where there exist an area integral and a regular or singular invariant measure.
Note that in the space of system
parameters these cases define, generally speaking, different regions which have a nonempty intersection.

\begin{itemize}
\item[{\bf 1.}] The case where there is no external field $\boldsymbol{b}=0$ (i.e., a balanced rigid body). This case is a generalization
of the well-known Euler case in the Euler-Poisson equations. In this case, the system~\eqref{Biz131} possesses a (singular)
invariant measure \cite{bkm1}
$$
\mu=(I_{13}\omega_1 + I_{23}\omega_2)^{-1}d\omega_1 d\omega_2 d\gamma_1 d\gamma_2 d\gamma_3.
$$

\item[{\bf 2.}] The case where the vector $\boldsymbol e$ defining the
constraint~\eqref{Eq:SuslovConstraint} is directed along the principal axis of inertia of the body and the system~\eqref{Biz131}
admits a~standard invariant measure \cite{Kozlov}
$$
\mu=d\omega_1 d\omega_2 d\gamma_1 d\gamma_2 d\gamma_3.
$$

\item[{\bf 3.}] The area integral $F_2=(\boldsymbol \gamma, {\bf I}
    \boldsymbol \omega)$ exists if the constraint vector $\boldsymbol
    e$ is\linebreak perpendicular to the circular section of the
    ellipsoid of inertia. In this case, the parameters of the
    system~\eqref{Biz131} satisfy the relations \cite{bkm}
\begin{equation*}
\begin{gathered}
I_{13}=0, \quad I_{23}^2 - I_{22}(I_{11} - I_{22})=0, \\
b_1=0, \quad I_{22}b_2 + I_{23}b_3=0.
\end{gathered}
\end{equation*}
We note that for $f_2=0$ the distribution given by this integral and by the constraint
\eqref{Biz03} is integrable and the system itself is holonomic. This fact was first established in \cite{Kozlov}.
\end{itemize}

In the general case (when there are no tensor invariants) the behavior of
the system~\eqref{Biz131} is typical of dissipative and nonholonomic
systems, i.e., the system exhibits multistability and strange
attractors~\cite{bkm}. Let us consider all the three cases in more detail.

\medskip

{\bf 1. Case $\boldsymbol b=0$}
The equations governing the evolution of the angular velocities $\omega_1$ and $\omega_2$ decouple in the system~\eqref{Biz131}:
\begin{equation}
\label{biz132}
\begin{gathered}
I_{11}\dot{\omega}_1=-\omega_2(I_{13}\omega_1 + I_{23}\omega_2), \quad
I_{22}\dot{\omega}_2=\omega_1(I_{13}\omega_1 + I_{23}\omega_2).
\end{gathered}
\end{equation}
They admit the energy integral
$$
E=\frac{1}{2}(I_{11}\omega_1^2+I_{22}\omega_2^2).
$$

\begin{figure}[!ht]
\centering
\includegraphics[totalheight=4.5cm]{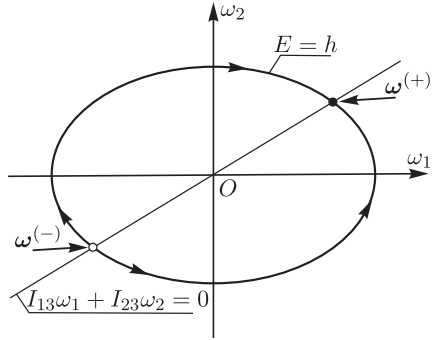}
\caption{A typical phase portrait of the system ($I_{13}>0$, $I_{23}<0$).}
\label{fig02}
\end{figure}

In this case, according to~\eqref{Biz131}, on the plane $\mathbb{R}^2=\{(\omega_1,
\omega_2)\}$ we have the straight line
$$
I_{13}\omega_1+I_{23}\omega_2=0,
$$
which is filled with fixed points. Thus,
\begin{itemize}
\item[{}] \it{in the subsystem~\eqref{biz132} each fixed level set of the energy integral $E=h$ with $h\ne 0$
consists of four trajectories: one stable fixed point and one unstable
fixed point and
a pair of ellipse arcs joining them (see Fig.~\ref{fig02}}).
\end{itemize}

In this case, its general solution (different from the fixed points)\\
is expressed explicitly in terms of exponential functions of time
\begin{equation*}
\begin{gathered}
\omega_1=\omega_0\frac{\sqrt{I_{11}}I_{22}}{I_{22}I_{13}^2+I_{11}I_{23}^2}
\frac{2\sqrt{I_{22}}I_{13}-\sqrt{I_{11}}I_{23}(e^u-e^{-u})}{e^u+e^{-u}},\\
\omega_2=\omega_0\frac{I_{11}\sqrt{I_{22}}}{I_{22}I_{13}^2+I_{11}I_{23}^2}
\frac{2\sqrt{I_{11}}I_{23}+\sqrt{I_{22}}I_{13}(e^u-e^{-u})}{e^u+e^{-u}},\\
u=\omega_0(t-t_0),
\end{gathered}
\end{equation*}
where $\omega_0$~is some constant which on the level set of the energy integral $E=h$
satisfies the relation
$$
\omega^2_0=\frac{2(I_{22}I_{13}^2+I_{11}I_{23}^2)}{I^2_{11}I^2_{22}}h.
$$

The fixed points of the subsystem~\eqref{biz132} in the system~\eqref{Biz131}
correspond to limit cycles (stable and unstable, respectively), their projections onto the Poisson sphere lie in the planes
perpendicular to the same vector
$$
\boldsymbol \xi=\left(\frac{I_{23}}{\sqrt{I_{13}^2+I_{23}^2}}, -\frac{I_{13}}{\sqrt{I_{13}^2+I_{23}^2}}, 0 \right).
$$

That is, on each level set of the energy integral $E=h$ we have two families of periodic solutions: one of them, $S_+$,
corresponds to a stable point of the system~\eqref{Biz131}, the other, $S_-$, corresponds to an unstable point. Each of the
families is parameterized by
$$
\Gamma=(\boldsymbol \xi, \boldsymbol \gamma)\in (-1, 1).
$$
The values $\Gamma=\pm 1$ correspond to the fixed points of the system~\eqref{Biz131}, which on the level set of the energy integral
$E=h$ are given by
\begin{equation}
\label{biz138}
\omega_1^{(\pm)}=\mp \sqrt{\frac{2h}{I_{11}I_{23}^2+I_{22}I_{13}^2}}I_{23}, \quad
\omega_2^{(\pm)}=\pm \sqrt{\frac{2h}{I_{11}I_{23}^2+I_{23}I_{13}^2}}I_{13},
\end{equation}
$$
\boldsymbol \gamma^{(\pm)}=\pm \boldsymbol \xi.
$$

\begin{figure}[!ht]
\begin{center}
\includegraphics[totalheight=5cm]{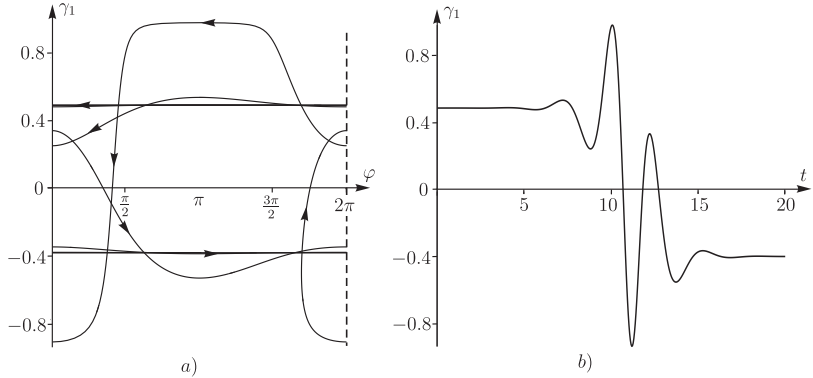}
\end{center}
\caption{Graphs of $\gamma_1(\varphi)$ and $\gamma_1(t)$ for one trajectory with the fixed moments of inertia
$I_{11}=2.44$, $I_{22}=1$, $I_{13}=0$, $I_{23}=0.5$ and the initial conditions
$\omega_1(0)=2$, $\omega_2(0)=10^{-4}$, $\gamma_1(0)=0.5$, $\gamma_2(0)=0.26$, $\gamma_2(0)=0.83$.}
\label{fig0112}
\end{figure}

All the other trajectories of the system~\eqref{Biz131}~are asymptotic: as
$t\to -\infty$ and $t\to +\infty$, they tend to one of the periodic
solutions of the families $S_-$ and~$S_+$, respectively (or to the fixed
points $\Gamma=\pm 1$), see Fig.~\ref{fig0112}.

In~\cite{bkm1}, cases were found where the system~\eqref{Biz131} (with zero potential) admits another first integral, and the
elements of the inertia tensor satisfy the relations
\begin{equation}
\label{eq834}
\begin{gathered}
1) \quad I_{13}=0, \quad I_{11}I_{22}=I^2_{22}+k^2I_{23}^2,\\
2) \quad I_{23}=0, \quad I_{11}I_{22}=I^2_{22}+k^2I_{13}^2,
\end{gathered}\qquad
 k=2n-1, \,\, n \in \mathbb{N}
\end{equation}
We note that the second case is obtained from the first by permutation of subscripts $1 \leftrightarrow 2$. Therefore,
we consider only the first case.

The additional integral turns out to be a polynomial homogeneous in $\boldsymbol \omega$,
$\boldsymbol \gamma$, of degree 1 in $\boldsymbol \gamma$ and degree $k$ in
$\boldsymbol \omega$. In particular, the first two integrals have the form
\begin{equation*}
\begin{array}{l}
n=1, \quad F_2=(\boldsymbol \gamma, {\bf I}\boldsymbol \omega)=I_{11}\gamma_1\omega_1+(I_{22}\gamma_2+I_{23}\gamma_3)\omega_2 \\ \\
n=2, \quad F_2=I_{11}^2\omega^2_1\big((I^2_{22}+I^2_{23})\gamma_1\omega_1+I_{22}(I_{22}\gamma_2+I_{23}\gamma_3)\omega_2\big)+\\
\qquad \qquad +I_{22}\omega^2_2\big(I_{22}(I^2_{22}+5I^2_{23})\gamma_2\omega_2+(I^2_{22}-3I^2_{23})(I_{11}\gamma_1\omega_{11}+I_{23}\gamma_3\omega_2\big)
\end{array}
\end{equation*}
For arbitrary $k$ the recurrent formula for the integral is presented in~\cite{bkm1} (see also \cite{Fedorov2009}).

Suppose that in this case the corresponding two-dimensional
integral\linebreak submanifold of the system is
$$
\mathcal{M}^2_{h, f}=\{(\boldsymbol \omega, \gamma)| E=h, \, F_1=1,\, F_2=f\}.
$$
If we assume that the value of $f$ is not equal to the values of the integral $F_2$ at the fixed points~\eqref{biz138}
$$
F_2(\boldsymbol \omega^{(+)}, \boldsymbol \gamma^{(+)})=f^{+}, \quad
F_2(\boldsymbol \omega^{(-)}, \boldsymbol \gamma^{(-)})=f^{-},
$$
then the restriction of the vector field of the system to $\mathcal{M}_{h, f}^2$ vanishes nowhere. Consequently,
when $f \ne f^{+}$ and $f \ne f^{-}$, the submanifold $\mathcal{M}_{h, f}^2$ (in the general case, each connected component of $\mathcal{M}_{h, f}^2$)
is diffeomorphic to the two-dimensional torus.

On each torus $\mathcal{M}^2_{h, f}$, in turn, there are two
limit cycles
$$
\mathcal{C}^+_{h, f}=\{(\boldsymbol \omega, \boldsymbol \gamma)| \boldsymbol \omega=\boldsymbol \omega^{(+)}, \, \boldsymbol \gamma^2=1, \, F_2(\boldsymbol \omega^{(+)},
\boldsymbol \gamma)=f\}
$$
$$
\mathcal{C}^-_{h, f}=\{(\boldsymbol \omega, \boldsymbol \gamma)| \boldsymbol \omega=\boldsymbol \omega^{(-)}, \, \boldsymbol \gamma^2=1, \, F_2(\boldsymbol \omega^{(-)},
\boldsymbol \gamma)=f\},
$$
where $\boldsymbol \omega^{(+)}$ and $\boldsymbol \omega^{(-)}$ are given by~\eqref{biz138}. As $t \to -\infty$, all
other trajectories
tend to the unstable cycle $\mathcal{C}_{h, f}^-$, while as $t \to
+\infty$, they tend to the stable cycle $\mathcal{C}_{h, f}^+$.

Let us take some unstable cycle $\mathcal{C}^{(-)}_*$ in the family $S_-$ and consider a set of trajectories
$\mathcal{M}^{(-)}_*$, which tend to it as $t \to -\infty$. In the case where the system has an additional integral
$F_2(\boldsymbol \omega, \boldsymbol
\gamma)$, this set $\mathcal{M}^{(-)}_*$ possesses the following natural property.
\begin{propos}
If $\frac{\partial F_2}{\partial \boldsymbol \gamma} \neq 0$, then, as $t\to +\infty$, all trajectories from $\mathcal{M}^{(-)}_*$
tend to the same stable cycle $\mathcal{C}^{(+)}_*$.
\end{propos}
{\proof Since the cycle $\mathcal{C}^{(-)}_*$ is an invariant
set, it lies on some fixed level set of the integral $F_2(\boldsymbol \omega, \boldsymbol
\gamma)=f_*$, and the entire set $\mathcal{M}^{(-)}_*$ lies on the same level set.

As $t\to +\infty$, we have $\boldsymbol \omega \to \boldsymbol \omega^{(+)}$, hence, all trajectories tend to the curve
$$
\mathcal{C}^{+}_*=\{(\boldsymbol \omega, \boldsymbol \gamma)|\boldsymbol \omega=\boldsymbol \omega^{(+)}, \, \boldsymbol \gamma^2=1, \, F_2(\boldsymbol \omega^{(+)},
\boldsymbol \gamma)=f_*\}.
$$}

\begin{figure}[!ht]
\centering
\includegraphics[totalheight=11cm]{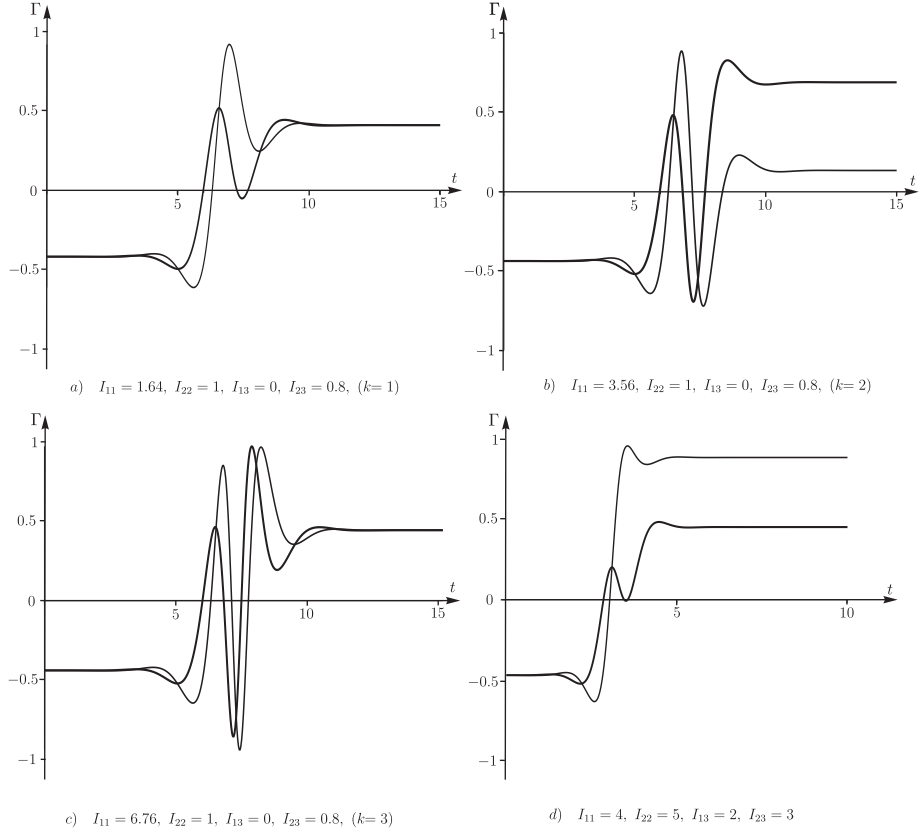}
\caption{Graph of $\Gamma(t)=(\boldsymbol \xi, \boldsymbol \gamma(t))$ for two trajectories started from the neighborhood
of the same unstable cycle, with different initial phases. Figures a -- c have been plotted for $I_{12}=0$ and the initial
conditions
$\omega_1(0)=2, \omega_2(0)=10^{-3}$, $\gamma_1(0)=-0.44$, $\gamma_2(0)=0.28$,
$\gamma_3(0)=0.85$ and $\omega_1(0)=2$, $\omega_2(0)=10^{-3}$, $\gamma_1(0)=-0.44$, $\gamma_2(0)=0.72$,
$\gamma_3(0)=0.53$. Figure d has been plotted for the initial conditions
$\omega_1(0)=3, \omega_2(0)=-2.001$, $\gamma_1(0)=-0.88$,
$\gamma_2(0)=-0.48$, $\gamma_3(0)=0$ and $\omega_1(0)=3$, $\omega_2(0)=-2.001$, $\gamma_1(0)=-0.49$, $\gamma_2(0)=0.26$,
$\gamma_3(0)=0.88$.}
\label{fig0222}
\end{figure}

As numerical experiments show (see Fig.~\ref{fig0222}), when conditions~\eqref{eq834} are not satisfied, different
trajectories from the set
$\mathcal{M}^{(-)}_*$ tend to different limit cycles as $t \to
+\infty$, hence
\begin{itemize}
\item[{}] {\it when $U=0$, the system~\eqref{Biz131} generally possesses no additional (analytical) integral.}
\end{itemize}

\begin{figure}[!ht]
\begin{center}
\includegraphics[totalheight=5cm]{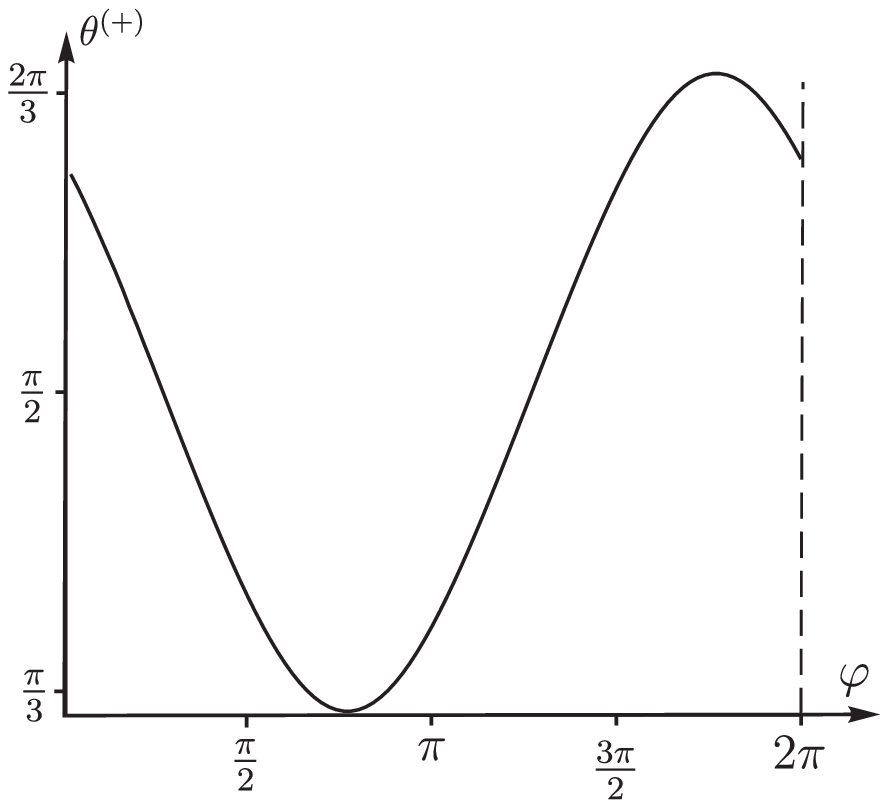}
\end{center}
\caption{Graph of $\theta^{(+)}(\varphi)$ for fixed $I_{11}=3.56$, $I_{22}=1$, $I_{13}=0$, $I_{23}=0.8$ and
$\omega_1(0)=2$, $\omega_2(0)=10^{-5}$, $\theta^{(-)}=\frac{\pi}{2}$.}
\label{fig0111}
\end{figure}

If we start a family of trajectories with different initial
azimuth angles (phases)
$\varphi_0$ in a neighborhood of the same unstable cycle specified on the Poisson sphere by the angle $\theta^-=\arccos \Gamma^{(-)}$, then, as $t \to +\infty$, we obtain a dependence of the angles $\theta^+(\varphi_0)=\arccos
\Gamma^+$ for those limit cycles to which the corresponding
trajectories tend (see Fig.~\ref{fig0111}).

\medskip

{\bf 2. Case $I_{13}=I_{23}=0$.} We now consider the case in which the vector $\boldsymbol{e}$ is directed along the principal
axis of the inertia tensor, i.e., $I_{13}=I_{23}=0$:

\begin{equation}
\begin{gathered}
\label{Biz135}
I_{11}\dot \omega_1=b_3\gamma_2-b_2\gamma_3, \quad I_{22}\dot \omega_2=b_1\gamma_3-b_3\gamma_1,\\
\dot \gamma_1=-\gamma_3\omega_2, \quad \dot \gamma_2=\gamma_3\omega_1, \quad \dot \gamma_3=\gamma_1\omega_2-\gamma_2\omega_1.
\end{gathered}
\end{equation}
As stated above, in this case the system~\eqref{Biz131} possesses a standard invariant measure.

Nevertheless, in the general case the system is Hamiltonian only after rescaling time.

\begin{theorem}
	The system~\eqref{Biz131} can be represented in the conformally Hamiltonian form
	$$
	\dot{x}_i=\gamma_3\{x_i, E(\boldsymbol x)\}, \quad E(\boldsymbol x)=\frac{1}{2}\big(I_{11}\omega^2_1+I_{22}\omega^2_2\big)+b_1\gamma_1+b_2\gamma_2+
b_3\gamma_3,
	$$
where $\boldsymbol x=(\omega_1, \omega_2, \gamma_1, \gamma_2, \gamma_3)$, and the nonzero Poisson brackets have the form
\begin{equation}
\begin{gathered}
\label{Biz133}
	\{ \omega_1, \gamma_2 \}=-\frac{1}{I_{11}}, \quad \{ \omega_2, \gamma_1 \}=\frac{1}{I_{22}}\\
	\{ \omega_1, \gamma_3 \}=\frac{1}{I_{11}} \frac{\gamma_2}{\gamma_3}, \quad \{ \omega_2, \gamma_3 \}=-\frac{1}{I_{22}}\frac{\gamma_1}{\gamma_3}.
\end{gathered}	
\end{equation}
\end{theorem}

The rank of the Poisson structure~\eqref{Biz133} is equal to 4, and the geometrical integral
$$
F_1=\gamma_1^2+\gamma_2^2+\gamma_3^2
$$
is a Casimir function of this structure. In addition, this conformally Hamiltonian representation is seen to have a singularity
when $\gamma_3=0$.\goodbreak

For the system~\eqref{Biz135} one can point out two more particular cases.
One of them can be trivially integrated, and was found by E.I.Kharlamova \cite{Char}, and the other generally admits
chaotic behavior
(see \cite{Kozlov}):
\begin{itemize}
\item[---] $b_3=0$, i.e., the radius vector of the center of mass lies in the plane orthogonal to the constraint
vector $\boldsymbol e$;
\item[---] $b_1=b_2=0$, i.e., the radius vector of the center of mass is collinear
with the constraint vector.
\end{itemize}
Let us consider them successively.

 $\boldsymbol b_3=0.$ In this case, equations~\eqref{Biz135} turn out to be invariant under the transformation
$$
\gamma_3\to -\gamma_3, \quad t\to -t.
$$
This yields the following natural result.
\begin{propos}
Let $\boldsymbol x(t)=(\omega_1(t), \omega_2(t), \gamma_1(t), \gamma_2(t),
\gamma_3(t))$~be a solution of the system~\eqref{Biz135} for $b_3=0$. Then
$
\boldsymbol {\tilde{x}}(t)=(\omega_1(-t), \omega_2(-t),
\gamma_1(-t), \\
\gamma_2(-t),
\gamma_3(-t))
$~is also a solution of this system.
\end{propos}

This observation allows us to divide all trajectories of the system into three types:
\begin{itemize}
\item[{--}] trajectories which never reach the equator
$(\gamma_3=0)$ on the Poisson sphere (each of them corresponds to the mirror image trajectory $\gamma_3 \to
-\gamma_3$, which is passed in the opposite direction);
\item[{--}] trajectories which transversally cross the equator $(\gamma_3=0)$ and which are periodic by virtue of the proposition;
\item[{--}] fixed points lying on the equator $(\gamma_3=0)$.
\end{itemize}

It follows that to analyze the behavior of the system trajectories, it suffices to consider only one half of the Poisson sphere.
For definiteness, we choose $\gamma_3>0$, make a change of variables and rescale time:
$$
\begin{gathered}
q_1=\sqrt{I_{11}}\gamma_2, \quad q_2=-\sqrt{I_{22}}\gamma_1, \quad p_1=\sqrt{I_{11}}\omega_1, \quad p_2=\sqrt{I_{22}}\omega_2,\\
\gamma_3 dt=d\tau.
\end{gathered}
$$
As a result, we obtain an integrable canonical Hamiltonian system with two degrees of freedom
(for more details on the Hamiltonization of nonholonomic systems, see \cite{BolsinovHam})
\begin{equation}
\begin{gathered}
\label{Biz136}
\frac{dq_i}{d\tau}=\frac{\partial H}{\partial p_i}, \quad
\frac{dp_i}{d\tau}=-\frac{\partial H}{\partial q_i}, \quad i=1, 2,\\
H=\frac{1}{2}\big(p^2_1+p^2_2\big)-\frac{b_2}{\sqrt{I_{11}}}q_1+\frac{b_1}{\sqrt{I_{22}}}q_2,
\end{gathered}
\end{equation}
which is defined inside the domain
\begin{equation}
\begin{gathered}
\label{Biz1360}
\frac{q_1^2}{I_{11}}+\frac{q_2^2}{I_{22}}\leqslant 1.
\end{gathered}
\end{equation}

The system \eqref{Biz136} describes the motion of a material point on the
plane under the action of a potential force. The trajectories of the point
are straight lines. Since the trajectory reaching the boundaries of
\eqref{Biz1360} is a half of the periodic trajectory (whose second half is
symmetrically reflected to the hemisphere\linebreak ${\gamma_3<0}$), we
find that in this case all trajectories (except for fixed points) are
periodic; the case of a quadratic potential is examined in \cite{Kozlov}.
\begin{remark}
This construction can obviously be generalized to the case of an arbitrary
potential field whose potential $U$ depends only on $\gamma_1$,
$\gamma_2$, in which case we obtain a natural Hamiltonian system in the
domain~\eqref{Biz136} with the Hamiltonian
\begin{equation}
	\label{eqs03}
	H=\frac{1}{2}(p_1^2 + p_2^2) + V(q_1, q_2), \quad V(q_1, q_2)=
U\left(-\frac{q_2}{\sqrt{I_{22}}}, \frac{q_1}{\sqrt{I_{11}}}\right)\!.
\end{equation}

In particular, this implies that the integrable potentials on the plane have their own integrable analogs in the Suslov problem.
The quadratic integrals~\cite{Okunaeva, Tatarinov1988}, as well as integrals of higher degrees on the plane \cite{Gramat},
can be carried over to the Suslov problem. However, when carried over to the Suslov problem, these cases must be topologically
modified
(taking into account the passage through the equator)
\end{remark}

 ${\boldsymbol b_1}={\boldsymbol b_2}=0.$ The system~\eqref{Biz135} also reduces to the problem of the motion of a material point
in a potential force field \cite{Kozlov}.
Indeed, we first fix the energy $E=h$ and express $\gamma_3$ as follows:
$$
\gamma_3=\frac{h}{2 b_3} - \frac{I_{11}\omega_1^2 + I_{22}\omega_2^2}{2 b_3}.
$$
Now, using this equation, we eliminate $\gamma_3$ in~\eqref{Biz135} and make a change of variables
(time rescaling is not required)
$$
\begin{gathered}
p_1=\gamma_1, \quad p_2=\gamma_2,\\
q_1=\frac{I_{22}}{b_3}\omega_2, \quad q_2=-\frac{I_{11}}{b_3}\omega_1.
\end{gathered}
$$
As a result, we obtain a natural system with two degrees of freedom with the canonical Poisson bracket
($\{ p_i,q_k \}=\delta_{ik}$) and the Hamiltonian
\begin{equation}
	\label{eqs02}
	H=\frac{1}{2}(p_1^2 + p_2^2) + \frac{1}{2}\left( \frac{h}{b_3} -
\frac{1}{2}\left( \frac{q_1^2}{I_{22}} + \frac{q_2^2}{I_{11}} \right)\right)^2.
\end{equation}
The system \eqref{eqs02} turns out to be integrable \cite{Kozlov, Mat} only in the case
$$
I_{11}=I_{22}.
$$

\medskip

{\bf 3. The existence of an area integral and isomorphism to the classical Hess case.}
\label{sec2_4}

In Section~\ref{sec2.1}, it was shown that when $U=0$, the
system~\eqref{Biz01} admits a~countable family of cases where there exists
an additional first integral. It turns out that the simplest of these
cases (when $k=1$) admits a natural generalization in the presence of a
gravitational field. This case is isomorphic to the classical Hess case,
which is treated in the Appendix. We recall the geometric meaning of the
corresponding restrictions on the parameters, assuming that all principal
moments of inertia are different:
\begin{itemize}
\item[{--}] make a transformation from the angular velocities to the angular momenta:
$$
\boldsymbol M={\bf I}\boldsymbol \omega;
$$
\item[{--}] consider in the three-dimensional space of angular momenta the level surface of the kinetic energy,
the {\it gyration ellipsoid}:
$$
(\boldsymbol M, {\bf I^{-1}}\boldsymbol M)=\const;
$$
\item[{--}]
the gyration ellipsoid possessing a pair of circular sections passes through the middle axis. In the Hess case
the center of mass lies on the perpendicular to the circular section of the gyration ellipsoid.
\end{itemize}
\begin{figure}[h!]
\centering
\includegraphics[totalheight=5cm]{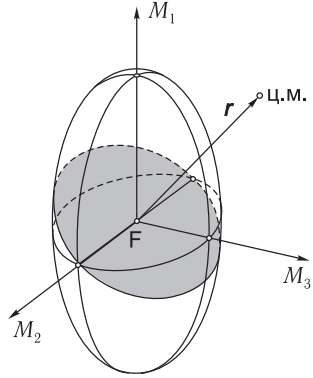}
\caption{Gyration ellipsoid and the location of the center of mass for the Hess case.}
\label{}
\end{figure}

Now let us rewrite the constraint equation~\eqref{Eq:SuslovConstraint} as
$$
({\bf I}\boldsymbol \omega, \boldsymbol a)=0, \quad \boldsymbol a={\bf I^{-1}\boldsymbol e}.
$$
It can be shown that if $Ox_1$~is the middle axis of inertia, then the condition for the vector $\boldsymbol n$
to be perpendicular to the circular section of the corresponding gyration ellipsoid coincides exactly with the conditions
for the existence of an integral with $U=0$ and $k=1$:
\begin{equation}
\label{bizi1}
I_{13}=0, \quad I_{11}I_{22}=I_{22}^2+I_{23}^2.
\end{equation}

In addition, we now assume that the center of mass of the body is also located on the normal to the same circular section in
the chosen coordinate system. This is equivalent to the conditions
\begin{equation}
\label{bizi2}
b_1=0, \quad I_{22}b_2+I_{23}b_3=0.
\end{equation}
We find that in this case the system~\eqref{Biz01} also admits the integral
$$
F=({\bf I}\boldsymbol \omega, \boldsymbol \gamma).
$$

Now we introduce the dimensionless parameters
$$
a=\frac{I_{23}}{I_{22}}, \quad b=-\frac{mgb_3}{\sqrt{I_{22}^2 + I_{23}^2}}
$$
and define the new variables $\boldsymbol{x}=(w_1, w_2, n_1,n_2,n_3)$:
\begin{equation}
\label{eq02}
\begin{gathered}
w_1=\omega_1, \quad w_2=\frac{\omega_1}{\sqrt{1 + a^2}}, \\
n_1=\gamma_1, \quad n_2=\frac{\gamma_2 + a\gamma_3}{\sqrt{1 + a^2}}, \quad
n_3=-\frac{a\gamma_2 - \gamma_3}{\sqrt{1 + a^2}}.
\end{gathered}
\end{equation}

The transformation \eqref{eq02} is a rotation of the coordinate system $Ox_1x_2x_3$ about the axis $Ox_1$ through the
angle ${\rm arctg} a$.
As a result, the coordinates of the center of mass take the form
$$\left( b_1, \frac{I_{22}b_2 + I_{23}b_3}{\sqrt{I_{22}^2 + I_{23}^2}}, \frac{I_{22}b_3 - I_{23}b_3}{\sqrt{I_{22}^2 + I_{23}^2}}\right)\!. $$
Thus, in the new coordinate system, in view of \eqref{bizi2}, the center of mass has been displaced only along the third axis.

The equations of motion \eqref{Biz131} in terms of the new variables become
\begin{equation}
\label{eq01}
\begin{gathered}
\dot{w}_1=-aw_2^2 - bn_2, \quad \dot{w}_2=aw_1w_2 + bn_1, \\
\dot{n}_1=-(an_2 + n_3)w_2, \quad \dot{n}_2=aw_2n_1 + w_1n_3, \quad
\dot{n}_3=n_1w_2 - n_2w_1.
\end{gathered}
\end{equation}

\begin{remark}
The system \eqref{eq01} is isomorphic up to a change of parameters to the Euler-Poisson equations on the invariant Hess relation
(see the Appendix).	
\end{remark}	
\begin{remark}
 In the case $b=0$ the system \eqref{eq01} possesses the particular solution
	$$
	\begin{gathered}
	w_1={\rm const}, \quad w_2=0, \\
 n_1=\cos\theta_0, \quad n_2=\sin(w_1t+\varphi_0)\sin\theta_0 \quad n_3=\cos(w_1t+\varphi_0)\sin\theta_0, \\
	\theta_0={\rm const}, \quad 	\varphi_0={\rm const}
	\end{gathered}
	$$
\end{remark}

The first integrals of the system \eqref{eq01} can be represented as
\begin{equation}
\label{eq03}
\begin{gathered}
E=\frac{1}{2}(w_1^2 + w_2^2) - bn_3, \quad  F_2=w_1n_1 + w_2n_2 \\
F_1=n_1^2 + n_2^2 + n_3^2=1.
\end{gathered}
\end{equation}

\begin{remark}
We note that equations \eqref{eq03} are also integrals of the vector field $\boldsymbol{u}$ (related to a rotation about
the symmetry axis)
\cite{RED}:
	$$
	\boldsymbol{u}=-w_2\frac{\partial }{\partial w_1} + w_1\frac{\partial }{\partial w_2}
	-n_2\frac{\partial }{\partial n_1} + n_1\frac{\partial }{\partial n_2}.
	$$
Nevertheless, $\boldsymbol{u}$ is not a symmetry field of the system \eqref{eq01}.
Indeed, if we denote the vector field of the system \eqref{eq01} by $\boldsymbol{v}$, we obtain
	$$
	[\boldsymbol{u}, \boldsymbol{v}]=aw_1\boldsymbol{u}.
	$$
\end{remark}

Let us examine in more detail the dynamics on the two-dimensional integral manifolds
$$
\mathcal{M}_{h,f}^2=\{\boldsymbol{x} \ |  \ E(\boldsymbol{x})=h, \ F_1(\boldsymbol{x})=1, \ F_2(\boldsymbol{x})=f \}.
$$
To do this, we parameterize them using the coordinates $(n_3, \varphi)$, where $\varphi$ is the angle variable:
$$
w_1=\sqrt{2(h + bn_3)}\sin\varphi, \quad w_2=\sqrt{2(h + bn_3)}\cos\varphi.
$$
Without loss of generality we set $b=1$.
Further, using \eqref{eq01}, we obtain the equations of motion in the form
\begin{equation}
\label{eq0003}
\dot{n}_3^2=2(h + n_3)(1 - n_3^2)-f^2, \quad \dot{\varphi}=-\frac{f}{2(h + n_3)} -
a\cos\varphi\sqrt{2(h + n_3)}.
\end{equation}

As can be seen, the gyroscopic function (defined by the equation for $n_3$)
coincides with the gyroscopic function for a spherical pendulum. Thus, we arrive at the well-known result for the Hess case.
\begin{propos}
The integrals of the system \eqref{eq01} become dependent in the following
cases:

1) the values of the first integrals lie on the curve given by the equation
\begin{equation}
\label{Geq3}
h=\frac{(1 - 3n_3^3)}{2n_3}, \quad f=\pm\frac{1 - n_3^2}{\sqrt{n_3}}, \quad n_3\in(0, 1),
\end{equation}

2) at the points
\begin{equation}
\label{Geq4}
f=0, \quad h=1 \quad \mbox{и} \quad f=0, \quad h=-1.
\end{equation}
\end{propos}

\begin{figure}[h!]
\centering
	\includegraphics[totalheight=5.5cm]{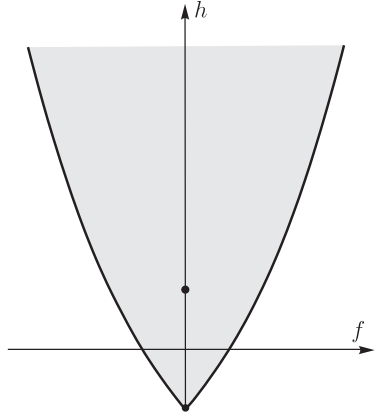}
	\caption{Bifurcation diagram (the grey area indicates the range of possible values of the first integrals).}
	\label{fig04}
\end{figure}

The resulting bifurcation diagram is shown in Fig. \ref{fig04}.
It can be shown that for values of the first integrals $h$ and $f$ which do not satisfy \eqref{Geq3} and \eqref{Geq4}
the level surface of the first integrals of the system \eqref{eq01} is diffeomorphic to the two-dimensional torus $\mathbb{T}^2$
whose vector field is described by the system \eqref{eq0003}. This system exhibits limit cycles, which is in good agreement with
the results of \cite{Kozlov1978, Ziglin}.

\begin{remark}
On the zero level set of the area integral $f=0$, at the
points~\eqref{Geq4} the system \eqref{eq0003} has singularities. Hence,
this case requires a separate\linebreak consideration.
\end{remark}

\begin{figure}[!ht]
	\centering
	\includegraphics[totalheight=6cm]{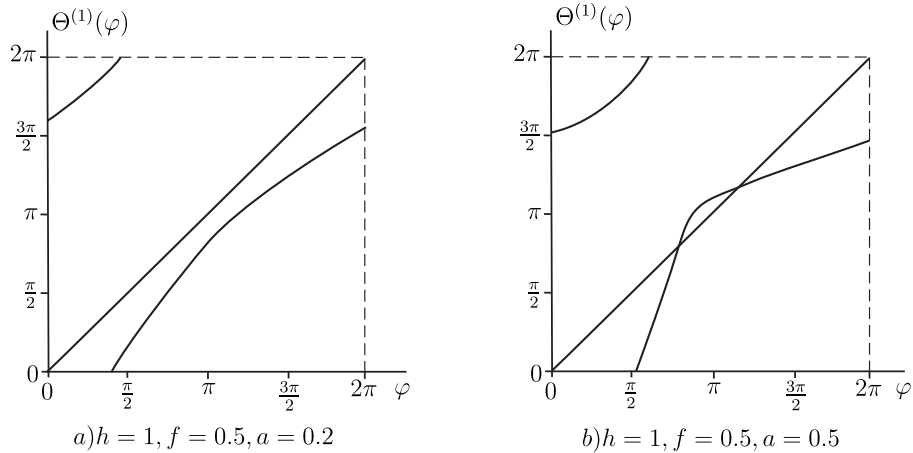}
	\caption{The map $\Theta^{(1)}(\varphi)$ and the diagonal of the square.}
	\label{fig_07}
\end{figure}

In the system \eqref{eq0003}
$n_3 \in (n_3^{(1)}, n_3^{(2)})$, where  $n_3^{(1)}$ and $n_3^{(2)}$
are solutions of the cubic equation $2(h + n_3)(1 - n_3^2)-f^2=0$.
In order to examine the vector field of \eqref{eq0003}, we consider on the torus $\mathbb{T}^2$ a Poincar\'{e} section
formed by the intersection with the plane $n_3=n_3^{(1)}$, which defines the map of the circle to itself:
$$
\Theta(\varphi): S^1 \to S^1.
$$
Figure \ref{fig_07} shows this map for different fixed parameters and the
diagonal (of the square), and the periodic solutions of \eqref{eq0003}
correspond to their intersection points. For the parameters corresponding
to Fig. \ref{fig_07}b we have two limit cycles: a~stable cycle and an
unstable one. Then, as the parameter $a$ decreases, the cycles disappear.

\clearpage
\section{A Chaplygin ball with Suslov's constraint.}

The paper~\cite{bkm1} is concerned with a system that is equivalent to the problem of the motion of a Chaplygin ball
with the additional Suslov constraint~\eqref{Eq:SuslovConstraint}. Moreover, this paper proposes
an implementation of this system which allows one to construct another possible nonholonomic generalization of the Euler-Poisson
equations (see Fig.~\ref{fig04}).

In this case, as in the implementation of Vagner~\cite{Vagner41}, it is
assumed that the rigid body $\mathcal{B}$ is equipped with wheels (on one
axis) and enclosed in a~fixed spherical shell. The condition that there be
no slipping in the direction\linebreak perpendicular to the plane of the
wheels
leads to the Suslov constraint:
$$
(\boldsymbol \omega, \boldsymbol e)=0,
$$
where $\boldsymbol \omega$~is the angular velocity of the body and $\boldsymbol e$~is the body-fixed vector lying in the plane
of the wheels perpendicularly to the axle supporting the wheels. Below we make use of a body-fixed coordinate system in which
$$
\boldsymbol e=(0, 0, 1).
$$
In addition, the body has a spherical cavity of radius $R$ whose center $O$ lies on the straight line joining the wheels.
At the point $P$
the cavity is in contact with a freely rotating homogeneous ball
$\mathcal{S}$ whose center is fixed. At the contact point $P$, the no-slip condition (mutual spinning is not prohibited)
is satisfied:
$$
R\boldsymbol \omega \times \boldsymbol \gamma=R_s \boldsymbol \omega_s\times \boldsymbol \gamma,
$$
where $\boldsymbol \omega_s$~is the angular velocity of the ball and $\boldsymbol
\gamma$~is the unit vector directed along the axis joining the centers of the cavity and the ball.

If in this implementation we choose the fixed ball
$\mathcal{S}$ inside the cavity in such a way that the vector joining its center
with the center of the body's cavity is vertical, then the equations of motion of the body $\mathcal{B}$ in the body-fixed
coordinate system take the form
\begin{equation}\label{eq29}
\begin{gathered}
\widehat{\mathbf I}\begin{pmatrix}
\dot\omega_1\\\dot\omega_2
\end{pmatrix}=\begin{pmatrix}
-(I_{13}\omega_1+I_{23}\omega_2)\omega_2+b_3\gamma_2-b_2\gamma_3\\
(I_{13}\omega_1+I_{23}\omega_2)\omega_1+b_1\gamma_3-b_3\gamma_1
\end{pmatrix}, \quad \omega_3=0\\
\dot{\gamma_1}=-\gamma_3\omega_2, \quad \dot{\gamma_2}=\gamma_3\omega_1, \quad
\dot{\gamma_3}=\gamma_1\omega_2-\gamma_2 \omega_1,\\
\widehat{\mathbf I}=\begin{pmatrix}
I_{11}+\mathcal{D}(\gamma_2^2+\gamma_3^2)&-\mathcal{D}\gamma_1\gamma_2\\
-\mathcal{D}\gamma_1\gamma_2&I_{22}+\mathcal{D}(\gamma_1^2+\gamma_3^2)
\end{pmatrix}, \quad \mathcal{D}=\frac{R^2}{R^2_s}I_s,
\end{gathered}
\end{equation}
where $I_s$~is the moment of inertia of the ball $\mathcal{S}$, $I_{ij}$~are the components of the body's tensor of inertia
relative to the point $O$, and the axes of the coordinate system have been chosen in such a way that~$I_{12}=0$.

\begin{figure}[!ht]
	\centering
	\includegraphics[totalheight=5cm]{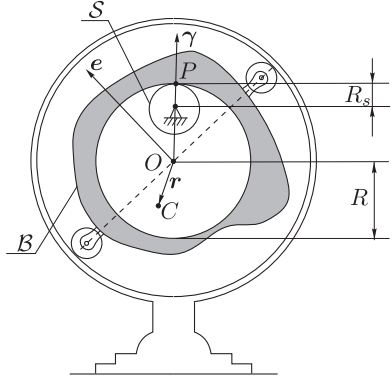}
	\caption{Realization of the system describing the dynamics of a Chaplygin ball with Suslov's constraint.}\label{fig05}
\end{figure}

Equations~\eqref{eq29} possess obvious integrals of motion: an energy integral and a geometrical integral:
\begin{equation}\label{eq30}
\begin{gathered}
E=\frac{1}{2}(\widehat{\mathbf I}\widehat{\boldsymbol\omega},\,
\widehat{\boldsymbol\omega})+(\boldsymbol b, \boldsymbol \gamma), \quad
F_1=(\boldsymbol\gamma,\,\boldsymbol\gamma)=1,
\end{gathered}
\end{equation}
where~$\widehat{\boldsymbol\omega}=(\omega_1,\,\omega_2)$. As for the previous system~\eqref{Biz131}, for integrability of
the system~\eqref{eq29} by the Euler\,--\,Jacobi theorem, we need an additional integral and a smooth invariant measure.

{\bf Case $\boldsymbol b=0$.} This simplest case of the system (absence of an external field) was considered in~\cite{bkm1},
where the existence of a singular invariant measure of the form
$$
\mu=(I_{13}\omega_1+I_{23}\omega_2)^{-1}d\omega_1 d\omega_2 d\gamma_1 d\gamma_2 d \gamma_3
$$
was not noticed. In this case, the energy integral~\eqref{eq30} can be written as
$$
E=\frac{1}{2}\Big((I_{11}+\mathcal{D})\omega_1^2+(I_{22}+\mathcal{D})\omega^2_2\Big)-
\frac{1}{2}\mathcal{D}(\boldsymbol \omega, \boldsymbol \gamma)^2
$$

\begin{figure}[!ht]
	\centering
	\includegraphics{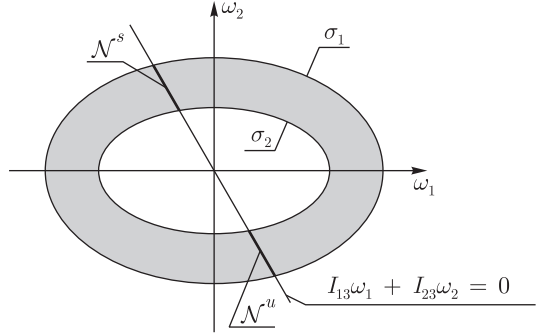}\vspace{-5mm}
	\caption{Region of possible motions for a fixed value of energy.\label{fig05Sy}}
\end{figure}

In the general case, the common level set of the first integrals~\eqref{eq30}
$$
\mathcal
M_h^3=\bigl\{\,\widehat{\boldsymbol\omega},\,{\boldsymbol\gamma}\mid E=h,\
\boldsymbol \gamma^2=1\,\bigr\}
$$
is a three-dimensional manifold which is projected onto the plane of angular velocities~$(\omega_1, \omega_2)$ into the strip
bounded by two ellipses (see Fig.~\ref{fig05Sy}):
$$
\sigma_1\colon 2h=I_{11}\omega_1^2 +I_{22}\omega_2^2,\qquad
\sigma_2\colon 2h=(I_{11}+mR^2)\omega_1^2+ (I_{22}+mR^2)\omega_2^2.
$$

As stated in Section~\ref{sec1.2}, the equation
$$
I_{13}\omega_1+I_{23}\omega_2=0
$$
defines an invariant submanifold $\mathcal{N}$ in the phase space of the system~\eqref{eq29}. Since $\widehat{\mathbf I}$
is positive definite, the angular velocities $\omega_1$ and $\omega_2$ on it remain constant. As can be seen from Fig.~\ref{fig05Sy},
the submanifold $\mathcal{N}$~is not connected and consists of two connected components (each of which is diffeomorphic to $S^2$)
$$
\mathcal{N}=\mathcal{N}^s \cup \mathcal{N}^u,
$$
one of which, $\mathcal{N}^s$,~is asymptotically stable and the other, $\mathcal{N}^u$,~is asymptotically unstable.
As shown in~\cite{bkm1}, the following result holds.

\begin{propos}
Each trajectory of the system \eqref{eq29} tends to $\mathcal N^s$ as $t\to+\infty$,
and to $\mathcal{N}^u$ as $t\to-\infty$.
\end{propos}

\begin{figure}[htb]
	\centering
	\includegraphics[totalheight=5cm]{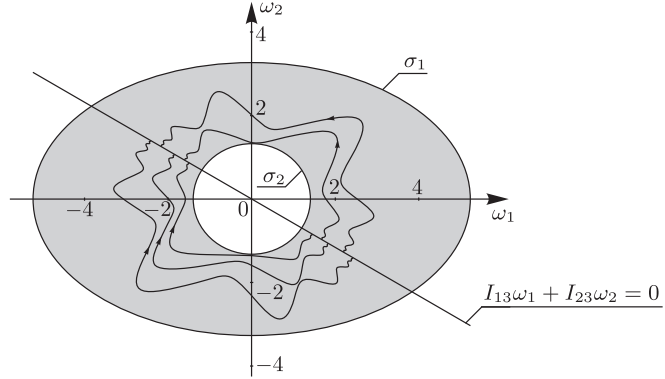}
	\caption{A typical view of projections of the trajectories onto the
plane~$(\omega_1,\,\omega_2)$ for~$I_{11}=1$, $I_{22}=1.5$, $I_{13}=0.7$,
$I_{23}=1.2$, $m=10$, $R=1$, $h=10$.}\label{fig6-sy}
\end{figure}

A typical view of projections of the system trajectories onto the plane~$(\omega_1,\,\omega_2)$ is shown in Fig.~\ref{fig6-sy}.

The invariant manifolds~$\mathcal N^u$ and $\mathcal N^s$ are filled with periodic trajectories for which the
vector~$\boldsymbol\omega$ is constant and the~vector~$\boldsymbol\gamma$ traces out circles on the sphere about the axis
given by the vector
$$
\boldsymbol \xi=\left(\frac{I_{23}}{\sqrt{I_{13}^2+I_{23}^2}}, -\frac{I_{13}}{\sqrt{I_{13}^2+I_{23}^2}}, 0\right).
$$

Thus, for each trajectory of the system~$\sigma(t)=\bigl(\boldsymbol\omega(t),\,\boldsymbol\gamma(t)\bigr)$
we have the limits~$\cos\theta_{{+}}=\lim\limits_{t\to+\infty}\left(\boldsymbol\gamma(t),\,\boldsymbol \xi\right)$ and~$\cos\theta_{{-}}
=\lim\limits_{t\to-\infty}\left(\boldsymbol\gamma(t),\,\boldsymbol \xi\right)$,
where~$\theta_{{+}}$ and $\theta_{{-}}$ are some constants.

{\bf Case $I_{13}=I_{23}=0$.} In this case the system~\eqref{Biz03} admits the standard invariant measure
$$
\mu=d \omega_1 d\omega_2 d\gamma_1 d\gamma_2 d\gamma_3.
$$
The question of the possibility of a Hamiltonian representation and integrable cases remains
open.

\clearpage
\section*{Appendix.\\
The Hess case in the Euler-Poisson equations}

This appendix is a shortened and revised version of a section of the book \cite{dtt}. However, this book is available
only in the Russian language and its results are unfortunately little known, although they would be useful to foreign
researchers.

The Euler\,--\,Poisson equations describing the motion of a heavy rigid body
with a fixed point have the following Hamiltonian form:
\begin{equation}
\label{hh01}
\dot{\boldsymbol{M}}=\boldsymbol{M}\times\frac{\partial H}{\partial \boldsymbol{M}} + \boldsymbol{\gamma}\times\frac{\partial H}{\partial \boldsymbol{\gamma}},
\quad
\dot{\boldsymbol{\gamma}}=\boldsymbol{\gamma}\times\frac{\partial H}{\partial \boldsymbol{M}}.
\end{equation}
Here the Hamiltonian $H$ is represented as
\begin{equation}
\label{hh001}
H=\frac{1}{2}(\boldsymbol{M}, {\bf A}\boldsymbol{M}) - \mu(\boldsymbol{r}, \boldsymbol{\gamma}),
\end{equation}
where $\boldsymbol{M}$~is the angular momentum vector in the coordinate system attached to the body, $\boldsymbol{\gamma}$~is the
unit vector of the vertical in the same system,
${\bf A}=\diag (a_1, a_2, a_3)$~is the inverse tensor of inertia, and $\boldsymbol{r}$~is the radius vector
of the center of mass of the body in the moving coordinate system.

Equations~\eqref{hh01} admit, in addition to the energy integral $H$, an area integral and a geometrical integral of the form
$$
F_1=(\boldsymbol{M}, \boldsymbol{\gamma}), \quad F_2=\boldsymbol{\gamma}^2=1.
$$

For the system \eqref{hh01} to be integrable in the sense of Liouville, we need another additional integral.
There are only a few known particular cases of integrability of equations~\eqref{hh01} in which this integral exists.
 All of them are realized under additional restrictions on the system parameters and on the initial conditions. These are
the cases of Euler, Lagrange, Kovalevskaya, and Goryachev\,--\,Chaplygin (see, e.g., \cite{dtt}). In the general case,
equations~\eqref{hh01} turn out to be nonintegrable.

The Hess case has the same number of free parameters (one parameter from the constants of the integrals disappears,
but an additional system parameter appears) as in the cases mentioned earlier and defines a family of particular solutions
given by the invariant relation
\begin{equation}
\label{g2p6f1}
	r_1M_1+r_3M_3=0,
\end{equation}
i.e., an isolated invariant manifold in the phase space.

The restrictions on the parameters in the Hess case have the form
\begin{equation}
\label{gess-*1}
\begin{gathered}
r_1\sqrt{a_3-a_2}\pm r_3\sqrt{a_2-a_1}=0,\\ r_2=0,
\end{gathered}
\end{equation}
and their physical meaning is the same as that described above in the
Suslov problem (see Section \ref{sec2_4}).

\begin{figure}[!ht]
	\centering
	\includegraphics{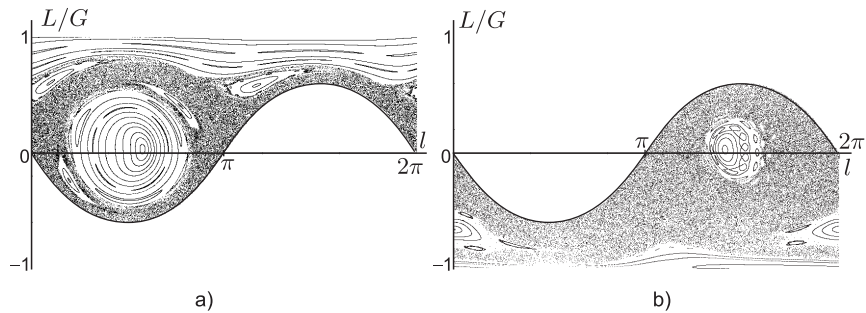}
\caption{Phase portrait (section formed by the intersection with the plane $g=\pi/2$)
for the Hess case under the conditions ${\bf I}=\diag(1,0.625,0.375)$, $\boldsymbol r=(3,0,4)$, $\mu=1.995$ with constant
integrals~$h=50.0$, $f=5.0$. Two stochastic layers divided by the double Hess separatrix are well visible:
points from one layer do not penetrate into the other. Figure~b also shows a {\it meandering torus} arising under
these conditions (see Fig.~\ref{gess2.pcx}).}
\label{gess1.pcx}
\end{figure}

\begin{figure}[!ht]
	\centering
	\includegraphics{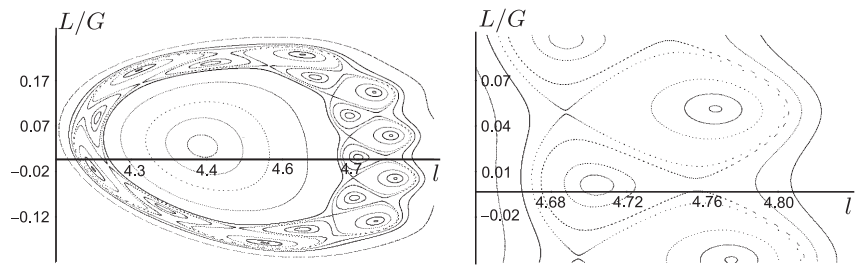}\caption{Meandering tori arising on the phase portrait in the Hess case
(the parameters are presented in Fig.~\ref{gess1.pcx})}
\label{gess2.pcx}
\end{figure}

 Generally speaking. the dynamics on this invariant Hess manifold differs from the usual quasi-periodic motion, which arises
when the conditions of the Liouville\,--\,Arnold theorem are satisfied.
Generally speaking, the Hess case cannot be integrated by quadratures
on~\eqref{g2p6f1}, but nevertheless can be analyzed\linebreak qualitatively.

\begin{remark}
In this section, we construct phase portraits by using the Andoyer\,--\,Deprit variables (see~\cite{dtt} for details).
\end{remark}

For certain values of the energy and area integrals the Hess relation can
define on the phase portrait a pair of double separatrices (see
Fig.~\ref{gess1.pcx}), which separate two chaotic zones (which show that
there exists no general integral under the Hess conditions). It is
interesting to note that in the phase space a~{\it meandering torus}
arises for the Hess case (see Fig.~\ref{gess2.pcx}). Such an effect is due
to the loss of twisting, and is encountered in Hill's celestial mechanics
problem~\cite{Simo, SS} and in the planar restricted three-body problem
\cite{Dullin}.

\begin{propos}
The Hess case in the Suslov problem (considered in Section~\ref{sec2_4})
is equivalent to the Hess case in the Euler\,--\,Poisson equations.
\end{propos}
{\proof Make a change of variables by which the Euler\,--\,Poisson
equations in the Hess case are reduced to \eqref{eq01}. To do this, we
explicitly write the Hamiltonian in the coordinate system for which one of
the axes, $Ox_3$, coincides with the axis perpendicular to the circular
section of the gyration ellipsoid:
\begin{equation}
\label{m15.9}
	H=\frac{1}{2}\left(a_1'(M_1^2+M_2^2)+a_3'M_3^2+2b'M_3M_1\right)-\mu\gamma_3,\quad \mu=\const.
\end{equation}
Such a coordinate system is no longer principal. The matrix of passage to new coordinates (from the system of principal axes)
can be expressed in terms of the components of the matrix $\bf A$ by the formulas
\begin{equation}
\label{m15.14}
	{\bf U}=\left( \begin{array}{ccc}
        \sqrt{\frac{a_3-a_2}{a_3-a_1}} & 0 & -\sqrt{\frac{a_2-a_1}{a_3-a_1}}\\
		0 & 1 & 0 \\
		\sqrt{\frac{a_2-a_1}{a_3-a_1}} & 0 & \sqrt{\frac{a_3-a_2}{a_3-a_1}}
	\end{array}\right).
\end{equation}
In this case, the invariant Hess relation~\eqref{g2p6f1} takes the form
\begin{equation}
\label{m15.10}
	M_3=0,
\end{equation}
and the equations of motion can be represented as
\begin{equation}
\label{gess05}
\begin{gathered}
\dot{M}_1=b'M_1M_2 - \mu\gamma_2, \quad \dot{M}_2=-bM_1^2+mu\gamma_1, \\
\dot{\gamma}_1=-a'_1 M_2\gamma_3 + b'M_1\gamma_2, \quad \dot{\gamma}_2=a'_1 M_1\gamma_3 - b'M_1\gamma_1, \\
\dot{\gamma}_3=a_1'(M_2\gamma_1 - M_2\gamma_1).
\end{gathered}
\end{equation}

After the change of variables
$$
\begin{gathered}
w_1=a_1'M_2, \ w_2=a_1'M_1, \\
n_1=-a_1'\gamma_2, \ n_2=-a_1'\gamma_1, \
n_3=a_1'\gamma_3
\end{gathered}
$$
and the change of parameters
$$
\mu=b, \ \frac{b'}{a_1'}=a,
$$
the system \eqref{gess05} is isomorphic to the system \eqref{eq01}.
}

To describe the motion of the rigid body in the fixed coordinate system, we introduce the variables $(\gamma_3,\varphi)$
\begin{equation}
\label{eq6-10}
\gamma_1=\sqrt{1-\gamma_3^2}\sin\varphi,\quad \gamma_2=\sqrt{1-\gamma_3^2}\cos\varphi,
\end{equation}
for which the equations of motion take the form
\begin{equation}
\begin{gathered}
\label{eq6-11}
	\dot\gamma_3=\pm\sqrt{2(1-\gamma_3^2)(h-U_{*})}, \\
	\dot\varphi=\frac{f\left(\gamma_3-b\sqrt{1-\gamma_3^2}\sin\varphi\right)}{1-\gamma_3^2}\mp
	b\cos\varphi\sqrt{2(h-U_{*})}
\end{gathered}
\end{equation}
where
$$
U_{*}=\frac{f^2}{2(1-\gamma_3^2)}-\mu\gamma_3, \quad H=h, \quad F_1=f
$$
The equation for the precession angle $\psi$ can be represented as
\begin{equation}
\label{eq6-12}
\dot\psi=\frac f{1-\gamma_3^2}.
\end{equation}

Using the quadratures for $\gamma_3$ and $\psi$,
N.\,E.\,Zhukovskii described the motion of the center of mass~\cite{Zhukovskij01}.
It is easy to see that it moves according to the law of a spherical pendulum. The solution for the angle of
proper rotation $\varphi$ cannot be obtained in terms of standard quadratures. Following\glossary{Некрасов}
P.\,A.\,Nekrasov~\cite{Nekrasov}, one usually reduces his definition to the
solution of an equation of Riccati type\index{Riccati!equation} (or to a linear equation with
doubly periodic coefficients).

Indeed, for the complex variable~$z=M_1+iM_2$ it is easy to obtain
$$
e^{-i\varphi}=-\frac{\dot\gamma_3+if}{\sqrt{1-\gamma_3^2}}\frac{z}{K^2},\quad
K^2=M_1^2+M_2^2=2(h-\mu\gamma_3),
$$
which for~$z$ leads to the nonlinear first-order equation
\begin{equation}
\label{klas-gess-2}
\dot z+\frac{ia_{13}}{2}z^2+\mu \frac{\dot\gamma_3+if}{K^2}z+
\frac12 i a_{13}K=0.
\end{equation}

In the case $f=0$ the system~\eqref{eq6-11} simplifies to give
\begin{equation}
\begin{gathered}
\label{eq6-14}
\dot\gamma_3=\sqrt{2(1-\gamma_3^2)(h+\mu\gamma_3)},\\
	\dot\varphi=\mp b\cos\varphi\sqrt{2(h+\mu\gamma_3)},
\quad \left(\text{или} \ \dot\theta=-\sqrt{2(h+\mu\gamma_3)}\right).
\end{gathered}
\end{equation}

In~\cite{Zhukovskij01}, Zhukovskii\glossary{Жуковский} showed that on the zero level set of the area integral
the trajectory of motion of the middle axis of the gyration ellipsoid forms at each instant of time a constant angle~$\theta$
(nutation angle) with the plane of the circular section
\begin{equation}
\label{klas-gess-3}
\sin\theta=\frac{a_2}{\sqrt{a_2(a_1+a_3)-a_1a_3}}.
\end{equation}
Using this result, it can be shown that on the zero level set of the area integral the middle axis of inertia moves along a
loxodrome.\index{Loxodrome}
In view of this characteristic motion Zhukovskii introduced the name {\em
	loxodromic pendulum}\index{Loxodromic!pendulum}
(of Hess), obtained practical conditions for such a motion
and proposed a mechanical model for its observation~\cite{Zhukovskij01}.

\begin{figure}[!htb]
    \centering
\includegraphics{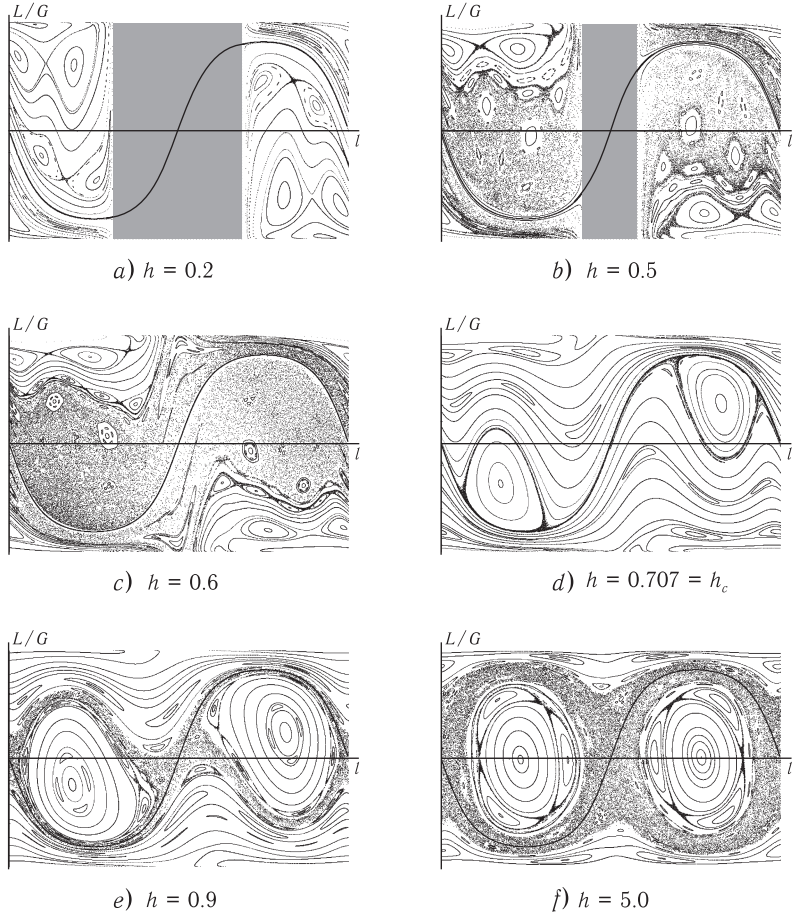}
	\caption{Phase portrait under the Hess conditions on the zero level set of the area integral
		$\left(H=\frac12\left(M_1^2+\frac23M_2^2+\frac12M_3^2\right)+\frac{1}{\sqrt3}\gamma_1+
		\frac{1}{\sqrt6}\gamma_3,\tilde{\mu}=h_c\right)$. It is evident from the figures that the torus corresponding to
the Hess integral at small energies is located in a regular foliation. The grey area indicates a~physically impossible range of values
of the variables.}
\label{gess-ris1}
\end{figure}

Let us consider the case of a loxodromic pendulum ($f=0$) in more detail (see Fig.~\ref{gess-ris1}).
From \eqref{eq6-14} we find
\begin{equation}
	\label{klas-gess-4}
	\dot\gamma_3^2 = 2(h-\tilde{\mu} \gamma_3)(1-\gamma_3^2),\quad
	\dot\psi=0,\quad \ln\Bigl(\tg\frac{l}{2}\Bigr)=\pm a_{13}K,
\end{equation}
where $M_1=K\sin l$, $M_2=K\cos l$.\goodbreak

\begin{figure}[!htb]
    \centering
\includegraphics{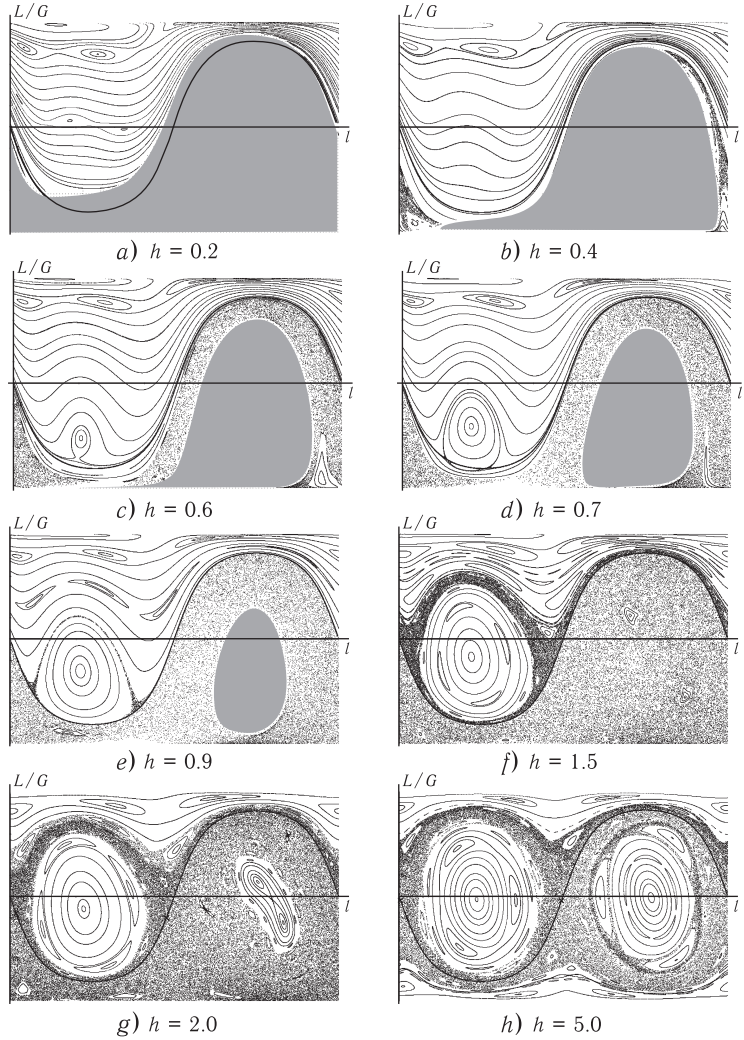}
	\caption{Phase portrait under the Hess conditions on the nonzero level set of the area integral~$f=1$
		$\left(H=\frac{1}{2}\left(M_1^2+\frac{2}{3}M_3^2+\frac{1}{2}M_3^2\right)+\frac{1}{\sqrt3}\gamma_1+
		\frac{1}{\sqrt6}\gamma_3\right)$. As above, at large~$h$
		the Hess solution separates two stochastic layers, and at small $h$ it lies in a regular foliation.}
\label{gess-ris2}
\end{figure}

There are two qualitatively different cases (this result was first obtained in the book \cite{dtt}):

\begin{list}{}{%
		\itemsep=1pt\parsep=0pt\leftmargin=2em\listparindent=2em\itemindent=0.5em
	}
	\item[\boldmath $h>\tilde{\mu}$.]
	The center of mass rotates in the principal circle (since $\psi=\const$). In this case the middle axis moves along the entire
loxodrome.
	In this case, on the phase portrait (Fig.~\ref{gess-ris1}\,e,~\ref{gess-ris1}\,f), which also contains chaotic trajectories,
the Hess solution separates two ``immiscible'' stochastic layers (see also Fig.~\ref{gess1.pcx}).
	The actual Hess solution in this case is not implementable: due to the instability the trajectory ``falls
down'' into one of these layers.
	
	As $h\to\infty$ (or $\tilde{\mu}\to 0$), everything reduces to the standard Euler case and the Hess solution tends to
the separatrix of permanent	rotation about the middle axis~\cite{KozlovMethods}.
	
	\item[\boldmath $h<\tilde{\mu}$.]
	The center of mass executes flat oscillations according to the law of a~physical pendulum, and the middle axis moves
according to~\eqref{klas-gess-3} along a segment of the loxodrome.
	In this case, the solution is periodic in absolute space (i.e., like the Goryachev solution, it is a one-frequency solution).
On the phase portrait (see Figs.~\ref{gess-ris1}\,a--\ref{gess-ris1}\,c) the Hess relation defines an invariant curve that
is entirely filled with fixed points and is located inside a regular foliation.
	
	For $f\ne 0$ the investigation of the motion is much more complicated and cannot be carried out analytically.
Figure~\ref{gess-ris2} shows a series of phase portraits illustrating the effect of divergence of the stochastic layers
(as energy~$h$ decreases) near the Hess solution, which becomes stable.
	
	In this case, the dynamics of absolute motion for small energies is
	three-frequency dynamics; as energy increases, the motion in one variable is asymptotic, and only two frequencies remain.\vspace{-3mm}
\end{list}

\begin{remark}
	As mentioned above, if one considers the perturbations of the
Euler\,--\,Poinsot problem
	under the Hess conditions, it turns out that the pair of separatrices emanating from unstable permanent rotations does not split
under a perturbation~\cite{KozlovMethods, Ziglin, Kozlov1978} (see Figs.~\ref{gess-ris1}\,f,
	\ref{gess-ris2}\,h). In this case, the integral~\eqref{g2p6f1} defines a singular torus filled with doubly asymptotic
trajectories approaching some unstable periodic solutions which,
as~$\tilde{\mu}\to0$, turn into permanent rotations about the middle axis. Such a description of the dynamics of a reduced system
does not
contradict the result obtained by Zhukovskii on  the quasi-periodic motion of the body's center of mass~\cite{Zhukovskij01}, since
	the system describing the motion of the center of mass is obtained by eliminating not the precession
angle, but the angle of proper rotation about the axis perpendicular to the circular section.
\end{remark}

{\bf Historical and critical comments.} Hess obtained his integral
when\linebreak searching for singular solutions of his own form of the
Euler\,--\,Poisson\linebreak equations~\cite{Hess} (1890), in which the
direction cosines have been eliminated using the integrals of motion. The
Hess case can be obtained from analysis of the branching of the general
solution on the complex plane of time.  This solution was overlooked by
Kovalevskaya~\cite{175} and arose for the first time in the work of
Appelrot \cite{175} (1892). However, as Appelrot himself wrote in the
original version of this work, he had made an error and overlooked this
case too. His oversight was pointed out by P.\,Nekrasov. In \cite{NN}
(1892) Nekrasov presented both the Hess conditions and the Hess integral
and reduced its integration to the Riccati equation. A more detailed
analysis shows that in the Hess case the solution branches out on the
complex plane of time (Appelrot, Lyapunov). The link between complex
branching, separatrix splitting and integrability was discovered by
S.\,L.\,Ziglin. In this vein he explained the enigmatic appearance of this
case. From the viewpoint of quasi-homogeneous systems and the
Kovalevskaya\linebreak exponents the Hess case is discussed in the recent
paper \cite{Kozlov20155}.

As mentioned previously, the geometrical analysis and the modeling of the
Hess top were proposed by Zhukovskii~\cite{Zhukovskij01}, and a detailed
analytical memoir on an explicit solution was written by
Nekrasov~\cite{Nekrasov} (1896).\glossary{Некрасов} The Hess integral was
independently rediscovered by Roger Liouville \cite{b2204-24} (1895), as
was the reduction to the Riccati equation (by the way, his note in Comp.
Rend. Acad. Sc. was submitted to А.\,Poincar\'{e}). In the next paper,
Liouville noted that the case found by him had been found earlier by Hess
(it was N.\,E.\,Zhukovskii who drew his attention to this fact) and
discussed the Maxwell principle, which is inapplicable in mechanics
problems.

In~\cite{b2204-25}, S.\,A.\,Chaplygin showed that the Hess motion can be
obtained for any body under the condition that the principal central
moments of inertia are different from each other. A link between the
invariant Hess relations and a pair of unsplit separatrices of the
perturbed Euler\,--\,Poinsot problem was established by
V.\,V.\,Kozlov~\cite{KozlovMethods} (see also \cite{Ziglin}). In the
Kirchhoff equations, an analog of the Hess case was noticed by
Chaplygin~\cite{chaplygin1}\glossary{Chaplygin} (who immediately used the
nonprincipal axes), and an identical case was obtained from the condition
of separatrix splitting in~\cite{kozlon}. For the problem of sliding of a
rigid body whose sharp edge is in contact with a smooth plane, an analog
of the Hess integral was found by G.\,V.\,Kolosov \cite{Koll} and
A.\,M.\,Lyapunov (who did not publish this result). An analog of the Hess
integral for other mechanics problems was found in \cite{Borisov2003,
BizyaevGess}. In \cite{Borisov2003}, its explicit symmetry origin is
elucidated for a wide class of potential systems. Various multidimensional
generalizations of the Hess case are discussed in \cite{DR01, DR02}.

We also mention the recent papers by P.\,Lubowiecki and
H.\,Zoladek~\cite{Lubowiecki2012, Lubowiecki2012II} (2012) and
A.\,V.\,Belyaev~\cite{Belaev2015} (2015). We note that the
paper~\cite{Belaev2015} contains results similar to those
of~\cite{Lubowiecki2012}, however, in all probability, the author,
although he referred to this paper, did not try to fully understand its
implications. The paper~\cite{Belaev2015} also contains very strange
asymptotic expansions the meaning of which for the dynamics is completely
unclear.

The basic theorems in~\cite{Lubowiecki2012} and~\cite{Belaev2015} are the
same, but in our opinion their proofs contain gaps. We consider in more
detail the paper~\cite{Lubowiecki2012}. There are no doubts about the
existence of two limit cycles which can merge at certain values of the
energy and area integrals. This fact is also pointed out in an extensive
memoir of Nekrasov~\cite{Nekrasov}, whose results have not been analyzed
from a modern point of view (see also \cite{Mlodz}). The conclusion that
further changes in the parameters (after bifurcation through the only
cycle on the torus) should lead to the appearance of tori with
quasi-periodic dynamics is unjustified.

The calculation of the real rotation number using complex linear equations
with doubly periodic coefficients is not correct and does not allow one to
obtain a~dependence on the system parameters. The result derived from the
theorem thus ``proved'' concerning the existence of a continuous invariant
measure\linebreak (without singularities) is also incorrect and is not
confirmed by further analysis. As is well known, the quasi-periodicity of
motion on the tori is a consequence of the existence of a smooth invariant
measure and the Diophantine properties of the Poincar\'{e} rotation
numbers (this constitutes the content of the well-known Kolmogorov
theorem~\cite{Kolmogorov}.

We note that in the above-mentioned papers \cite{Lubowiecki2012,
Belaev2015} the conclusion is drawn that at a rational rotation number the
torus is foliated by degenerate periodic trajectories. This phenomenon is
typical of Hamiltonian systems (with a smooth invariant measure). However,
in the Hess case the system has no smooth invariant measure, and it is
well known that, for systems without an invariant measure, the graph of
the rotation number versus a first integral is a Cantor ladder. For the
nonholonomic problem of the rolling motion of a ball (with a displaced
center of mass) this effect was discovered in \cite{BolsinovMer, Bizzz}.
Another example of vector field  on tori (related to the Hill's equation)
for which the rotation number depending on the system parameters
represents the Cantor ladder is given in~\cite{Simo}. The horizontal
segments of the Cantor ladder (which correspond to the rational rotation
number) correspond to tori on which there are either one or several limit
cycles. That's why the following problem is open.

{\bf The open problem.} {\it Is the rotation number on tori (corresponding
to the Hess case of the Euler\,--\,Poisson equations) depending on system
parameters the Cantor manifold? Are there any limit cycles by the rational
rotation number?}

There is some confusion in~\cite{Lubowiecki2012, Lubowiecki2012II} regarding normal hyperbolicity.
In fact, the cycle is hyperbolic on the invariant Hess manifold; on the isoenergetic surface there can be no
normal hyperbolicity due to the existence of an invariant measure (induced by the standard volume form of the Euler-Poisson equations),
and under perturbations (deviations from the Hess conditions) the invariant surface is not preserved
and the splitting of separatrices to hyperbolic points leads to the formation of a stochastic layer \footnote{By the way,
all these effects (like the main conclusions from~\cite{Lubowiecki2012,
Lubowiecki2012II,Belaev2015})
were illustrated by a Poincar\'{e} map in~\cite{Borisov2003}
(as well as in the book~\cite{dtt}), to which no reference is made in the above-mentioned works.}.

The authors are grateful to V.V.Kozlov, who gave a careful reading to the
manuscript and made a number of important comments.

\end{document}